\documentclass[useAMS,usenatbib,usegraphicx]{mn2e}
\usepackage{txfonts}

\def\ts{\langle t_{\rm s} \rangle}
\def\MJ{M_{\rm J}}
\def\itp{i_{\rm tp}}
\def\ME{M_{\rm E}}
\def\RH{R_{\rm H}}
\def\yr{\;\rm yr}
\def\spose#1{\hbox to 0pt{#1\hss}}
\def\lta{\mathrel{\spose{\lower 3pt\hbox{$\sim$}}
    \raise 2.0pt\hbox{$<$}}}
\def\gta{\mathrel{\spose{\lower 3pt\hbox{$\sim$}}
    \raise 2.0pt\hbox{$>$}}}

\title[Planets and Asteroids around $\gamma$ Cephei]{Planets and
Asteroids in the $\gamma$ Cephei system}

\author[P. E. Verrier \& N. W. Evans]{P. E. Verrier\thanks{E-mail:
pverrier@ast.cam.ac.uk (PEV); nwe@ast.cam.ac.uk (NWE)} and
N. W. Evans\footnotemark[1]\\ Institute of Astronomy, University of
Cambridge, Madingley Road, Cambridge, CB3 0HA, United Kingdom}

\begin{document}

\label{start}
\maketitle

\begin{abstract}
  The binary star system $\gamma$ Cephei is unusual in that it
  harbours a stable giant planet around the larger star at a distance
  only about a tenth of that of the stellar separation. Numerical
  simulations are carried out into the stability of test particles in
  the system. This provides possible locations for additional planets
  and asteroids. To this end, the region interior to the planet is
  investigated in detail and found to permit structured belts of
  particles. The region between the planet and the secondary star
  however shows almost no stability. The existence of an
  Edgeworth-Kuiper belt analogue is found to be a possibility beyond
  $65$ au from the barycentre of the system, although it shows almost
  no structural features. Finally, the region around the secondary
  star is studied for the first time. Here, a zone of stability is
  seen out to $1.5$ au for a range of inclinations. In addition, a ten
  Jupiter-mass planet is shown to remain stable about this smaller
  star, with the habitability and observational properties of such an
  object being discussed.
\end{abstract}

\begin{keywords}
stars: individual: $\gamma$ Cephei (HR 8974) -- planetary systems --
binaries: general -- methods: \textit{N}-body simulations
\end{keywords}

%--------------------------------------------------------------

\section{Introduction}

$\gamma$ Cephei is one of the closest separation binary systems that
contains a planet. The primary and secondary stars are separated by
18.5 au. The planet (designated $\gamma$ Cep Ab) has a minimum mass of
1.7 $\MJ$ and follows an eccentric orbit about the larger star
alone. The observational parameters of $\gamma$ Cephei, as determined
by \citet{Ha03}, are summarised in~Table~\ref{tab:orbels}.

Although the data are reasonably reliable, there is some uncertainty
regarding the masses of the components. The inclination to the line of
sight of a spectroscopic binary is usually indeterminate. This means
that the mass of the primary is generally reckoned from spectral
type. Then, the mass function (e.g., Smart 1977) permits a minimum
mass to be assigned to the secondary, albeit crudely (Griffin,
Carquillat \& Ginestet 2002).

For $\gamma$ Cephei, the mass of the primary is $\approx 1.57
M_{\odot}$ from photospheric modelling \citep{Fu04}. The most recent
determination from stellar evolution models is $\approx 1.7
M_{\odot}$~\citep{Af05}. The mass of the secondary is given by
\citet{Dv03} as $\approx 0.4 M_{\odot}$. However, assuming the
original value of the primary's mass of 1.57 $M_{\odot}$, a minimum
mass can be derived as $\approx 0.34$ $M_{\odot}$ from the velocity
amplitude fitted by \citet{Ha03}.

There have been a number of studies of the dynamics of the $\gamma$
Cephei system to date. Foremost is the numerical investigation of
\citet{Dv03}, which does use an earlier, and slightly different, set
of orbital parameters (see Table \ref{tab:altorbels}). They used
Burlisch-Stoer and Fast Liapunov Indicator methods to show that the
$\gamma$ Cephei system is stable over Myr time-scales. They carried
out test particle integrations as a guide to the possible existence of
further planets.  The results show a stable inner region between 0.5
and 0.8 au and then an additional stable zone at low inclination
around 1 au. \citet{Dv03} noted that this coincides with the 3:1 mean
motion resonance. This resonance is stable in the $\gamma$ Cephei
system, but is unstable in the Solar system. \citet{Dv03} conclude
that Earth-mass planets (up to 90 $\ME$) can exist in this region
which is fortunately just on the edge of the habitable zone
\footnote{The habitable zone has boundaries which depend on
assumptions regarding stellar luminosity and effective temperature, as
well as the climate model adopted for the hypothetical habitable
planet.}. An extension to this work by \citet{Dv04} showed that the
planet's eccentricity is less important than that of the two stars for
the stability of a second planet.  \citet{Ha05} also carried out
numerical studies of the system's stability using a Burlisch-Stoer
integrator, for a range of possible configurations of the planet's and
binary's semimajor axes, eccentricities and inclinations, confirming
and extending the results of \citet{Dv03}.

Here, we use both numerical simulations and analytic calculations to
study zones of stability as possible locations of additional planetary
companions to either star and asteroid or Edgeworth-Kuiper belt
analogues.  Edgeworth-Kuiper belts are of particular interest in
binary systems, as they may be being observed (indirectly) in
exosystems such as $\tau$ Ceti and $\eta$ Corvi (Greaves et al. 2004;
Wyatt et al. 2005). The results fall into three categories: possible
planetary companions and asteroid belts centred on the primary star
(\S 2), planetary companions around the secondary star (\S 3) and
finally possible Edgeworth-Kuiper belt about both stars (\S 4).

\begin{table*}
\centerline{
\scriptsize
\begin{tabular}{|l|r@{ $\pm$ }l|r@{ $\pm$ }l|r@{ $\pm$ }l|} 
\hline
Name                                 &\multicolumn{2}{|c|}{$\gamma$
  Cep A}      & \multicolumn{2}{|c|}{$\gamma$ Cep B} &
\multicolumn{2}{|c|}{$\gamma$ Cep Ab } \\
\hline
Class                                &\multicolumn{2}{|c|}{K1IV sub-giant star} &  \multicolumn{2}{|c|}{M dwarf star } & \multicolumn{2}{c}{Planet}\\
Mass                                 & 1.59 & 0.12 $M_{\odot}$
&  \multicolumn{2}{|c|}{0.4 $M_{\odot}$} & 1.7        & 0.4 $M_{J}$\\
Period (days)                        &\multicolumn{2}{|c|}{--}
& 20750.6579 & 1568.6 & 905.574    & 3.08\\
Semimajor axis (au)                 &\multicolumn{2}{|c|}{--}
& 18.5       & 1.1 & 2.13       & 0.05\\
Eccentricity                         &\multicolumn{2}{|c|}{--}                  & 0.361      & 0.023 & 0.12       & 0.05\\
Longitude of periastron ($^{\circ}$) &\multicolumn{2}{|c|}{--}
& 158.76     & 1.2  & 49.6       & 25.6\\
Time of periastron passage (JD)      &\multicolumn{2}{|c|}{--}
& 2448429.03  & 27.0 & 2453121.925 & 66.9\\
\hline
\end{tabular}}
\large
\caption{Best fit orbital parameters for the $\gamma$ Cephei system
from \citet{Ha03}. Mass of star B is from \citet{Dv03}.}
\label{tab:orbels}
\end{table*}

\begin{table}
\centerline{
\scriptsize
\begin{tabular}{|l|c|c|c|} 
\hline
Name                 & $\gamma$ Cep A      & $\gamma$ Cep B & $\gamma$ Cep Ab\\
\hline
Class                & K1IV sub-giant star & M dwarf star   & Planet\\
Mass                 & 1.6$M_{\odot}$      & 0.4$M_{\odot}$ & 1.76$M_J$\\
Period (days)        & --                  & 25567.5        & 902.2 \\
Semimajor axis (au) & --                  & 21.36          & 2.15\\
Eccentricity         & --                  & 0.44           & 0.209\\
\hline
\end{tabular}}
\large
\caption{Orbital parameters for the $\gamma$ Cephei system used by
  \citet{Dv03}.}
\label{tab:altorbels}
\end{table}

%--------------------------------------------------------------

\section{Planets and Asteroid Belts around the Primary}

\subsection{Algorithm}
\label{sec:algorithm}
For all the simulations, we use the parameters of the $\gamma$ Cephei
system given in Table~\ref{tab:orbels}. Here and elsewhere in the
paper, the equations of motion are integrated using a conservative
Burlisch-Stoer method provided in the {\tt MERCURY} software package
\citep{Ch99}. Although not as fast as sympletic methods, this was
chosen because of its ability to provide close encounter data and
handle highly eccentric objects. The two stars and planet are
simulated as point masses for gravitational interactions. Any
additional objects are taken as massless test particles to decrease
integration times. Test particles are removed from the simulations
when they collide with the primary star, or pass an ejection distance
of the order of several hundred au.  A collision with the primary
means that the test particle has a position that lies within the body
of the primary, as judged by its stellar radius of $0.02$ au (Hatzes
et al. 2003). Close encounters are allowed to occur, and are defined
as taking place whenever a test particle enters within one Hill radius
$\RH$ of the secondary star or the planet, defined as
\begin{equation}
\RH = a \left( {\frac{m}{3M}} \right)^{1/3} 
\end{equation}
where $m$ is the mass of the secondary or planet and $M$ is the mass
of the primary. This works out as $\approx 8.1$ au for the secondary and
$\approx 0.15$ au for the planet.

To maintain accuracy, the variation in the system's total energy and
angular momentum is monitored throughout each simulation. Using this
to constrain the initial timestep to 1 day and the tolerance in the
Burlisch-Stoer algorithm \citep{Pr99} to $10^{-12}$ leads to an
overall fractional change in the system's energy $\Delta E/E$ of about
$10^{-8}$ over a 100 Myr period. This can therefore be considered the
maximum time the system can be accurately followed. All the
simulations presented here are typically 1 Myr in time-scale, for which
$\Delta E/E \approx 10^{-11}$ or better.

It is straightforward to show that the $\gamma$ Cephei system has long
term stability. A 100 Myr integration shows no major variation in the
orbits. Regular short period variations do occur, for example, a
slight oscillation of the planet's semimajor axis over time-scales
equal to both its orbital period and the binary's orbital period. An
additional secular variation is seen over a period of about 5500
years, evident in the eccentricity and longitude of the planet
only. This secular period is in good agreement with the results of
quadrupole theory (see eq. (36) of \cite{Fo00}).

%--------------------------------------------------------------
\begin{figure}
\includegraphics[width=\hsize]{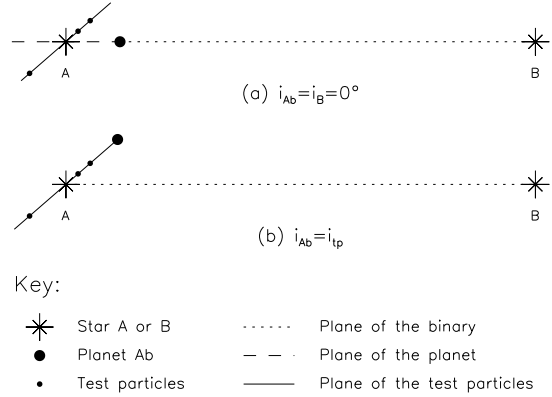}
\caption{\label{fig:inc_illus} Two possible configurations for test
particles. Top: the planet remains in the plane of the binary and the
test particles are inclined relative to this. In other words, the
inclination of the planet $i_{\rm Ab}$ is the same as the inclination
of the secondary $i_{\rm B}$ and both are zero. Bottom: the planet and
test particles have the same inclination relative to the plane of the
binary. In other words, $i_{\rm Ab}$ is equal to the inclination of
the test particles $\itp$.}
\end{figure}

\begin{figure}
\includegraphics[width=\hsize]{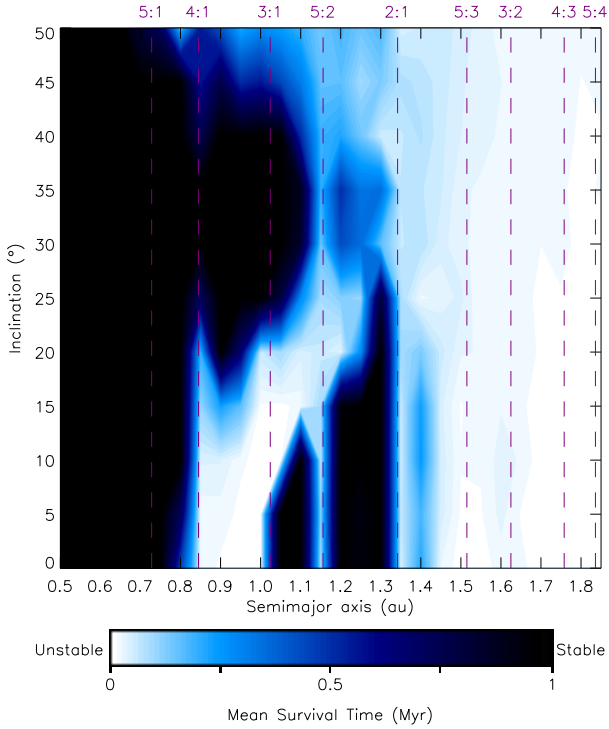}
\caption{\label{fig:stabmapinner} Stability map for test particles
interior to the orbit of the planet in the coplanar case $i_{Ab} =
0$. Here, stability is indicated by the test particle's survival time
averaged over longitudes and normalised to 1 Myr.  Darker colours
indicate more stable regions, whilst lighter colours show less stable
regions. Over-plotted in purple are the nominal locations of the mean
motion resonances with the planet Ab, up to the 5:4 case.}
\end{figure}

\subsection{Test Particles Interior to the Orbit of the Planet}

\label{sec:tpip}

Figure~\ref{fig:inc_illus} shows two possible configurations of test
particles in the system considered here. In the first, the test
particles are inclined to the common plane of the binary and
planet. In the second, both the test particles and planet are
coplanar, yet inclined relative to the plane of the binary.  The
binary is always assumed to be viewed edge-on (that is, it lies in a
plane perpendicular to the plane of the sky). If the planet is at an
inclination $i_{\rm Ab}$ relative to this, its mass must be increased
accordingly by dividing by $\sin i$ where $i = 90^\circ - i_{\rm Ab}$.

With the planet and binary coplanar, we begin by investigating the
stability of test particles in the region interior to the known
planet. A grid of particles with semimajor axis from 0.5 to 1.85 au
and inclination from 0$^{\circ}$ to 50$^{\circ}$ with resolution 0.05
au and 5$^{\circ}$ respectively is integrated for 1 Myr.  Thirty-six
particles are started at each grid point with varying initial
longitudes of pericentre ($\omega = 0^\circ, 60^\circ, \dots
300^\circ)$ and longitudes of ascending node ($\Omega = 0^\circ,
60^\circ, \dots 300^\circ$). The orbits are initially circular.  The
stability is then determined by computing the mean survival time $\ts$
in Myrs at each grid point averaged over the 36 test particles. The
results are shown in Figure~\ref{fig:stabmapinner} and can be compared
to figure 2 of \citet{Dv03}. Note that our stability index is slightly
different to the criteria used by \citet{Dv03}, who removed test
particles after they become orbit crossing. This may miss the
occasional test particle that is stable, for example, if it lies in a
Trojan-like orbit.  We only remove test particles if they collide with
the central star or are ejected from the system.

The map shows test particles with semimajor axes less than $\approx
1.4$ au are stable. However, there is a strip of instability between
roughly 0.8 and 1.0 au, creating an island of stability at low
inclinations between 1.0 and 1.3 au.  Some of the structure in the map
can be clearly related to the positions of the mean motion resonances
(MMRs) with the planet (indicated in Fig.~\ref{fig:stabmapinner}). The
4:1, 3:1, 5:2 and 2:1 resonances seem to mark transitions from
stability to instability.  For example, the 5:2 MMR divides the island
at $\approx 1.15$ au.  The lack of effect of some of the higher order
MMRs, such as the 5:1 case, is probably due to their comparative
weakness and narrowness. The lack of stability beyond $\approx 1.4$ au
is readily explained. The gravitational reach of the planet as a
multiple of the Hill radius (Jones, Underwood \& Sleep 2005) places
the limit on stability at 1.31 au, a good match with the results
here. Note that in calculating this limit the maximum
eccentricity of the planet obtained during the simulation has been
used.
\begin{figure}
\includegraphics[width=\hsize]{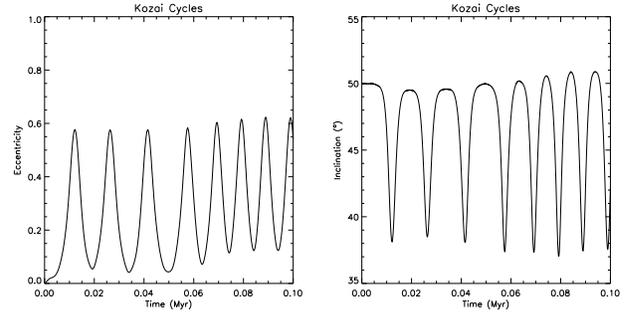}
\caption{\label{fig:orbitsb} The variation of eccentricity and
inclination with time for a test particle undergoing Kozai cycles. The
starting elements are $a = 0.5$ au and $i= 50^\circ$.}
\end{figure}
\begin{figure}
\includegraphics[width=\hsize]{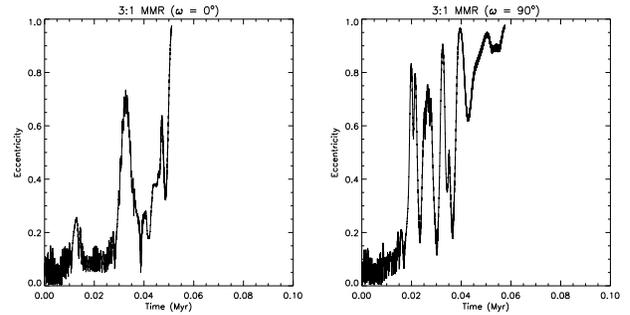}
\caption{\label{fig:orbitsa} The variation of eccentricity with time
for test particles at the 3:1 MMR for starting $\omega = 0^\circ$ and
$90^\circ$.}
\end{figure}
Many of the test particles in the high inclination region of
Figure~\ref{fig:stabmapinner} show evidence of Kozai cycles, as
illustrated by the orbits in Figure~\ref{fig:orbitsb}. The
\citet{Ko62} instability is well-known from studies of high
inclination comets and asteroids in the Solar system. It sets in at
inclinations greater than a critical value of $i_{\rm crit} = {\rm
asin} \sqrt{0.4} \approx 39.23^\circ$. During Kozai cycles, the
eccentricity and inclination vary so as to maintain approximate
constancy of the integral of motion $I_{\rm K} = \sqrt{1-e_{\rm tp}^2}
\cos i_{\rm tp}$, where $e_{\rm tp}$ is the test particle's
eccentricity. As the semimajor axis increases, the amplitude of
eccentricity and inclination librations becomes larger, thus
accounting for the increased instability evident in this region of
Figure~\ref{fig:stabmapinner}.

As already seen, the MMRs with the planet are important in shaping the
regions of stability. However, one major feature unexplained by this
is the instability strip between roughly 0.8 and 1.0 au. Although at
zero inclination the edges are marked by the 4:1 and 3:1 MMRs, the
instability strip shows a pronounced evolution with inclination which
is suggestive of a secular resonance instead. The classical
Laplace-Lagrange linear theory \citep{Mu00}, although derived for low
eccentricity and inclination regimes around a dominant central mass,
can be applied to give a first approximation of the locations of these
resonances.  A secular resonance for a test particle occurs when its
precession rate has exactly the same magnitude as an eigenfrequency of
the system.  The eigenfrequencies are easily calculable for the three
body system made up of the two stars and planet and are the
eigenvalues of the 2x2 matrices $\mathbf{A}$ and $\mathbf{B}$
respectively, which have components
\begin{eqnarray}
A_{jj} & = & + n_j \frac{1}{4} \frac{m_k}{M+m_j} \alpha \bar{\alpha} b^{(1)}_{3/2}(\alpha),\nonumber\\
A_{jk} & = & - n_j \frac{1}{4} \frac{m_k}{M+m_j}  \alpha \bar{\alpha} b^{(2)}_{3/2}(\alpha),\nonumber\\
B_{jj} & = & - A_{jj},\\
B_{jk} & = & - A_{jk} \frac{ b^{(1)}_{3/2}(\alpha)}{ b^{(2)}_{3/2}(\alpha)}.\nonumber
\end{eqnarray}
Here, $n_j$ is the mean motion of object $j$ (1 represents the planet
and 2 represents the secondary), $m_j$ and $a_j$ are the mass and
semimajor axis of object $j$, $M$ is the mass of the primary (the
central object), $\alpha = a_1/a_2$ and $b^{(1)}_{3/2}(\alpha)$ and
$b^{(2)}_{3/2}(\alpha)$ are Laplace coefficients. Using the values for
the masses and semimajor axes given in Table \ref{tab:orbels} along
with $n_1 = 145.2^\circ \yr^{-1}$, $n_2 =6.337 ^\circ \yr^{-1}$ and
$\alpha=0.1151$ gives the Laplace coefficients as
$b^{(1)}_{3/2}(\alpha)=0.3542^\circ \yr^{-1}$ and
$b^{(2)}_{3/2}(\alpha)=0.05089^\circ \yr^{-1}$, employing the {\tt
MATHEMATICA} routines of \citet{Mu00}. Calculating the matrices and
solving for the eigenfrequencies gives
\begin{eqnarray}
g_1 &=&0.04338^\circ \yr^{-1},\nonumber\\
g_2 &=& 0.00005211^\circ\yr^{-1},\\
f   &=& -0.04343^\circ\yr^{-1},\nonumber
\end{eqnarray}
where $g_1$ and $g_2$ are the eigenvalues of $\mathbf{A}$ and $f$ is
the degenerate eigenvalue of $\mathbf{B}$. Note that the $g_1$ and $f$
eigenfrequencies have about the same magnitude, whilst $g_2$ is almost
zero due to the large mass ratio between the planet and secondary
star. The precession rate of the test particle is given by
\begin{equation}
A_{\rm tp} = n \frac{1}{4} \left(\frac{m_1}{M}\alpha_1\bar{\alpha}_1b^{(1)}_{3/2}(\alpha_1) + \frac{m_2}{M}\alpha_2\bar{\alpha}_2b^{(1)}_{3/2}(\alpha_2) \right)
\end{equation}
where $n$ is the particle's mean motion. For the region interior to
the planet $\alpha_j = \bar{\alpha}_j= a/a_j$, where $a$ is the test
particles semimajor axis. Plotting $A_{\rm tp}$ as a function of $a$
from 0.5 to 1.85 au shows a resonant location where the $g_1$ and $f$
eigenfrequencies intersect the curve at $\approx 0.8$ au. This
supports the idea that the location of the inner edge of the
instability strip on the map coincides with a secular resonance.

Comparing our results with those of \citet{Dv03}, it is easy to see
that the broad trends are similar, despite slightly differing orbital
elements. The main stable regions are slightly closer to the star in
\citet{Dv03}. This may be due to a higher eccentricity of both the
planet and the secondary, which means that they approach the central
star more closely, reducing separations with the test particles.
\citet{Dv03} find that the 3:1 MMR is stable, in contrast to asteroids
in the Solar System in the same resonance with Jupiter. Here, we find
that the resonance is unstable, with a particle following a fairly
steady evolution until switching to a mode where its eccentricity is
rapidly driven to unity on a time-scale of 10 kyrs, as shown in
Figure~\ref{fig:orbitsa}.

\begin{figure}
\includegraphics[width=\hsize]{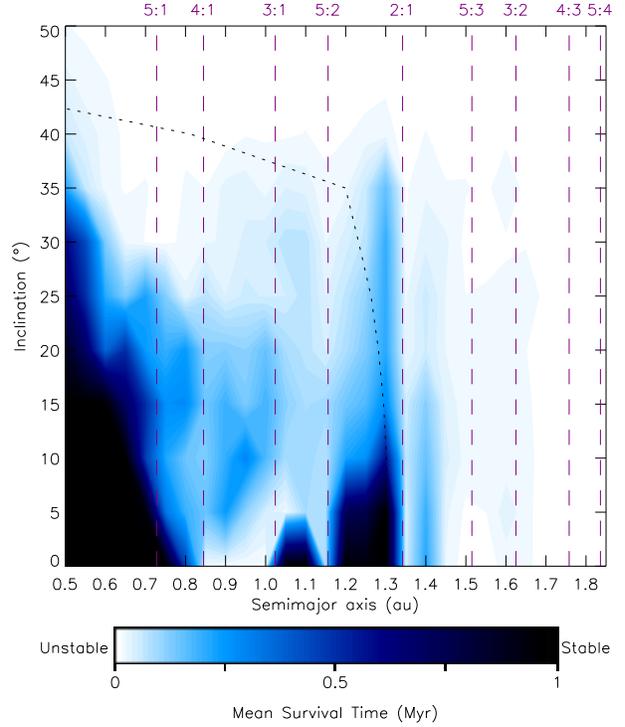}
\caption{\label{fig:stabmapincl} As Figure~\ref{fig:stabmapinner}, but
  here the planet and test particles are all inclined to the plane of
  the binary.  Over-plotted in purple are the nominal locations of the
  mean motion resonances with the planet Ab, up to the 5:4 case. The
  dotted curve shows the gravitational reach of the planet, calculated
  using the method of Jones et al. (2005).}
\end{figure}
%
%\begin{figure}
%\includegraphics[width=\hsize]{fig5.eps}
%\caption{\label{fig:planetKoz} 
%  {\bf CUT The variation of semimajor axis, eccentricity and inclination with
%  time for $\gamma$ Cephei Ab with starting inclination of $45^\circ$
%  to the orbital plane of the binary.}}
%\end{figure}

The case where the planet is also inclined ($i_{Ab} \neq 0$) has not
been previously studied. This is a more likely configuration for a
system that has formed in a common protoplanetary disc. Here, we
investigate this case by using the same grid of test particles as
before, but with the planet sharing the same inclination as the test
particles and with $\Omega_{Ab} = 0$. This means that, to reproduce
the same radial velocity dataset, the mass of the planet must be
increased by dividing by $\sin i$, where $i = 90^\circ - i_{Ab}$.

The results are displayed in Figure~\ref{fig:stabmapincl}. As compared
to the earlier case of Figure~\ref{fig:stabmapinner}, the unstable
region has expanded, especially at high inclinations. This is partly
caused by the change in the extent of the gravitational reach of the
planet, as shown by the dotted curve in
Figure~\ref{fig:stabmapincl}. Although this does scale with the
increasing planetary mass, the sharp change at $40^\circ$ inclination
is due to the planet becoming subject to Kozai cycles.  The large
increase in eccentricity here means that the planet's periastron is
much closer to the star, and hence its gravitational influence is
larger.  At $50^\circ$, the planet is unstable, colliding with the
central star after $\approx 0.5$ Myr.

At first sight, it may seem that the rest of the increased instability
evident in Figure~\ref{fig:stabmapincl} is caused by the increased
mass of the planet. However, experiments show that this is not so.
For example, an integration of the $30^\circ$ case with the planet
inclined but at minimum mass ($1.7 \MJ$) shows very little difference
to Figure~\ref{fig:stabmapincl}. This suggests that the cause lies in
the relative inclination of test particle and planet to the plane of
the binary. In the case of Figure~\ref{fig:stabmapinner}, the
amplitude of libration of $i_{\rm tp}$ and $e_{\rm tp}$ of test
particles is generally modest for all cases below $i_{\rm crit}$, the
critical value for the Kozai instability. For test particles in
Figure~\ref{fig:stabmapincl}, the amplitude is no longer small and
becomes larger with increasing inclination, rapidly driving particles
into the regime where the Kozai instability is effective. This is
understandable as in the former case, the forces due to the masses in
the system are always directed towards the plane of the binary,
whereas in the latter case this is not true. As the inclination
increases, the misalignment between the forces due to the stars and
the force due to the planet increases, and so the amplitude of
libration increases.

\subsection{Test Particles Interior to the Orbit of the Binary}

\label{sec:tpib}

Holman \& Wiegert (1999) have already studied the stability of test
particles in binary systems. These may orbit either a single star or
both stars. Here, we study test particles around the primary star.
For this case, Holman \& Wiegert (1999) introduced the notion of a
critical semimajor axis $a_{\rm crit}$, which is the largest circular
orbit around the primary for which a ring of test particles survives
for at least $10^4$ binary periods ($\approx 600$ kyrs in the case of
$\gamma$ Cephei).  Using their eq~(1), we find $a_{\rm crit} = 4.0 \pm
0.6$ au.  In other words, our expectation is that all test particles
starting out at semimajor axes greater than 4.0 au will be rapidly
swept out.  The existence of the comparatively large and eccentric
planet $\gamma$ Cephei Ab will cause further de-stabilization.

To investigate this, simulations of test particles in the region from
0.5 to 20.0 au are carried out. They have initially zero eccentricity
and are set up on a grid with resolution 0.5 au in starting semimajor
axis.  The range of inclinations is restricted from $0^\circ$ to
$30^\circ$ for the prograde case, and from $150^\circ$ to $180^\circ$
for the retrograde case, both in steps of $10^\circ$.  Seventy-two
particles are started at each grid point with varying initial
longitudes of pericentre ($\omega = 0^\circ, 30^\circ, \dots
330^\circ)$ and longitudes of ascending node ($\Omega = 0^\circ,
60^\circ, \dots 300^\circ$). The orbits of the test particles are
followed for 1 Myr, and any close encounters are recorded.  Although
this is a limited range of inclinations, it is expected that those not
investigated are largely unstable. This receives confirmation from
exploration integrations in the case of $70^\circ$ inclination, for which
very few test particles survive throughout the entire region.

\begin{figure}
\includegraphics[width=\hsize]{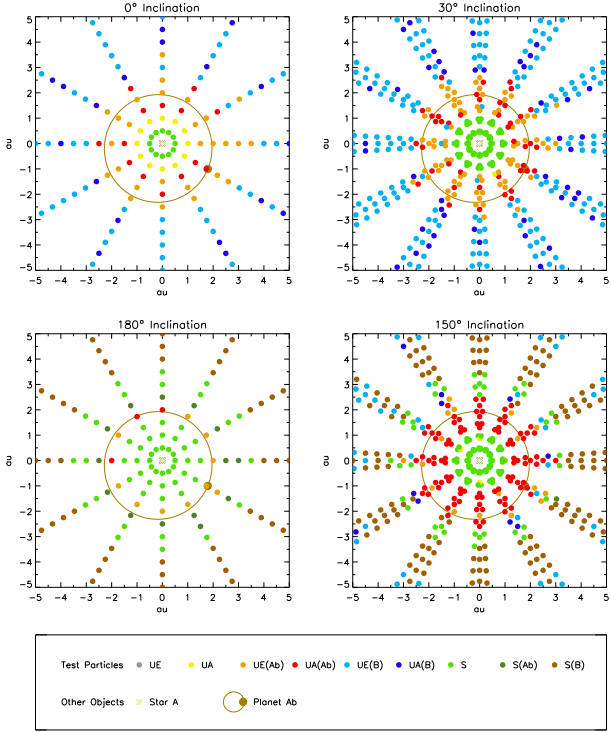}
\caption{\label{fig:inner} The ultimate fate of test particles
interior to the orbit of the binary, with inclinations $0^\circ,
30^\circ, 150^\circ$ and $180^\circ$. For clarity, only test particles
in and above the plane of the binary and planet are displayed. The
following abbreviations are used in the key: UE = unstable, ejected
from the system; UA = unstable, collided with star A; UE(Ab) =
unstable, ejected from the system after a close encounter with Ab;
UA(Ab) = unstable, collided with A after a close encounter with Ab;
UE(B) = unstable, ejected from the system after a close encounter with
B; UA(B) = unstable, collided with star A after a close encounter with
B; S = stable; S(Ab) = stable, but suffered a close encounter with Ab;
S(B) = stable, but suffered a close encounter with B.}
\end{figure}
\begin{figure}
\includegraphics[width=\hsize]{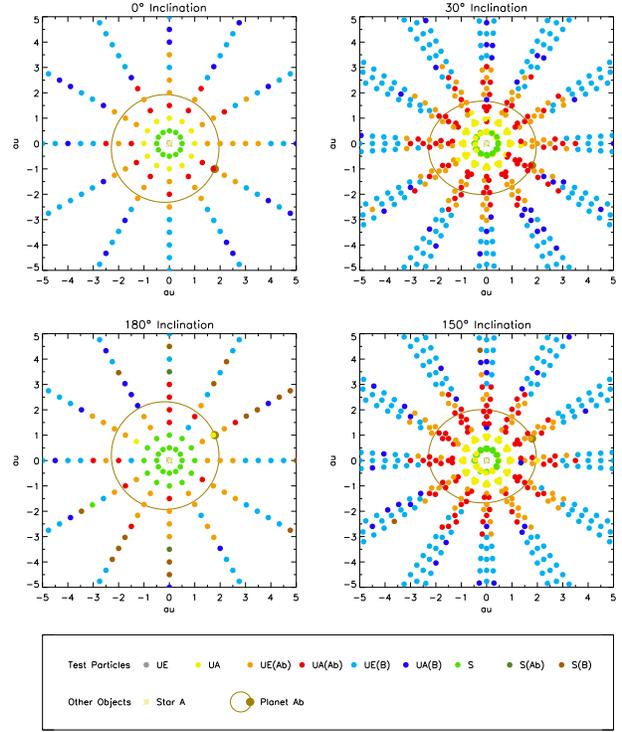}
\caption{\label{fig:innerAb} 
  As Figure~\ref{fig:inner}, but the test particles and the planet are
  similarly inclined.}
\end{figure}

Figures~\ref{fig:inner} and \ref{fig:innerAb} show the ultimate fates
of the test particles within 5 au. They differ in that the planet is
in the plane of the binary and test particles are inclined in
Figure~\ref{fig:inner}, whilst both planet and test particles are
similarly inclined in Figure~\ref{fig:innerAb}.  Four panels showing
the results of selected simulations are displayed in each case,
corresponding to prograde with $i_{\rm tp} =0^\circ$ and $30^\circ$,
retrograde with $i_{\rm tp} = 180^\circ$ and $150^\circ$. 

%The very inner regions ($0.5$ to $1.8$ au) show good agreement with
%Section~\ref{sec:tpip}. 
In the inner regions, test particles close to the planet are swept out
on a precession time-scale ($\approx 5.5$ kyrs). Within $\approx 3$
au, prograde test particles are either ejected or collide with the
central star after a close encounter with the planet. This is evident
from the particles coloured orange and red in the upper panels of both
Figures~\ref{fig:inner} and \ref{fig:innerAb}.  The retrograde case is
different. The lower panels of Figure~\ref{fig:inner} show swathes of
stable particles coloured either brown or green, according to whether
they reside within the Hill sphere of the secondary or not. Note that
the secondary's periastron is at $11.8$ au, so particles out to $3.7$
au are within its Hill sphere at some point. In the case $i_{\rm tp} =
180^\circ$, the retrograde stability zone extends out to $\approx 7$
au.  As the inclination decreases ($i_{\rm tp} \rightarrow
150^\circ$), the stability zone shrinks to the annuli between $0.5$ to
$1.0$ au and $3.0$ to $5.0$ au.  The lower panels of
Figure~\ref{fig:innerAb} show the case when both test particles and
planet are retrograde.  In the case $i_{\rm tp} = 180^\circ$, the
large stability region seen previously has almost completely
disappeared. There are only a few remaining test particles that
survive the 1 Myr integration.  The similarity of the two right-hand
panels of this figure shows that the inner region's evolution is
almost entirely controlled by the planet.

\begin{figure*}
\includegraphics[width=0.7\hsize]{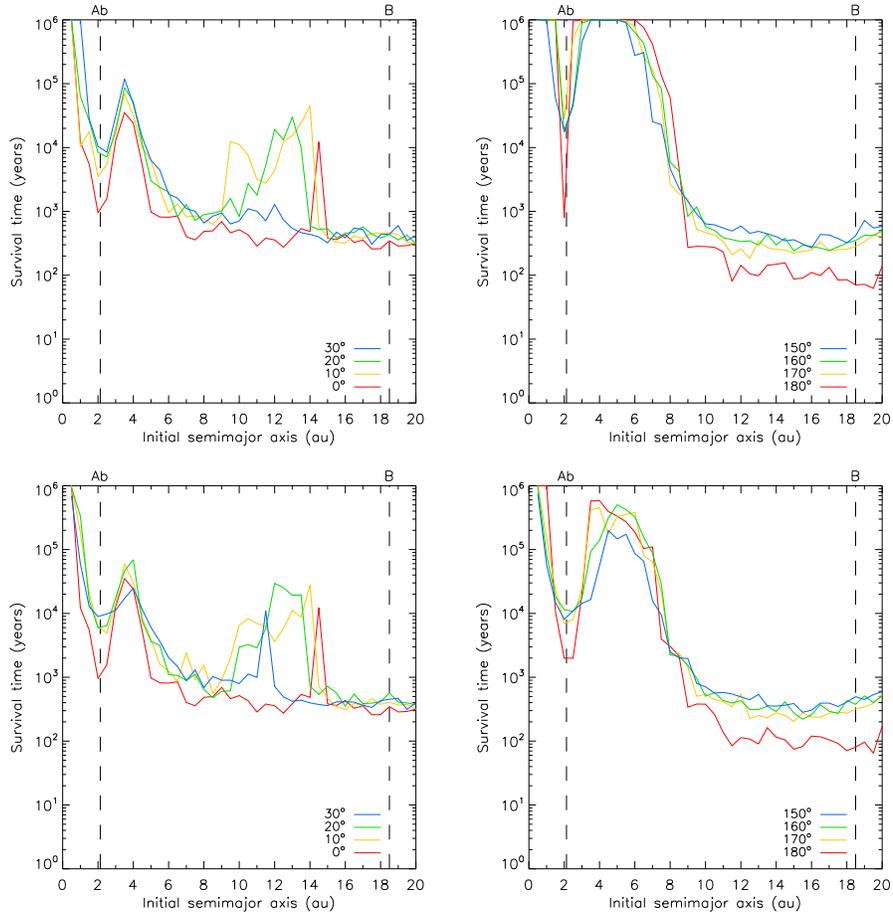} 
\caption{\label{fig:survivtimes} The mean survival time of test
  particles as a function of starting semimajor axis for particles
  interior to the orbit of the binary. In the left-hand panels,
  the starting inclination of the test particles is $0^\circ$ (red),
  $10^\circ$ (yellow), $20^\circ$ (green) and $30^\circ$ (blue); in
  the right-hand panels, it is $180^\circ$ (red), $170^\circ$
  (yellow), $160^\circ$ (green) and $150^\circ$ (blue).  The top
  panels are inclined test particles only, the bottom panels are
  similarly inclined test particles and planet.  The starting
  semimajor axis of the secondary star and the planet are shown as
  vertical dashed lines.}
\end{figure*}
\begin{table*}
\caption{\label{table:pevtable} The number of test particles interior 
  to the orbit of the binary at each starting semimajor axis and 
  inclination that survive for 1 Myrs. There are 72 test particles 
  initially for all cases except $i=0^\circ$ and $i=180^\circ$, which 
  have 12.
}
\centerline{ \scriptsize
\begin{tabular}[width=0.9\textwidth]{c||c|c|c|c|c|c|c|c|c|c|c|c|c|c|c||c|c|c|c|c} \hline
\null &\multicolumn{20}{c}{Starting Semimajor Axis [in au]}\\ 
Inclination & 0.5 & 1.0 & 1.5 & 2.0 & 2.5&3.0 & 3.5 & 4.0 & 4.5 & 5.0 & 5.5 & 6.0 & 6.5 & 7.0 & 7.5 & 12.0 & 12.5 & 13.0 & 13.5 & 14.0 \\
\hline
$0^\circ$   & 12 & - & - & - & - & - & - & - & - & - & - & - & - & - & - & - & - & - & - & - \\
$10^\circ$  & 72 & - & - & - & - & - & - & - & - & - & - & - & - & - & - & - & - & - & - & 2 \\
$20^\circ$  & 72 & - & - & - & - & - & - & - & - & - & - & - & - & - & - & 1 & - & 2 & - & - \\
$30^\circ$  & 72 &72 & - & - & - & - & - & - & - & - & - & - & - & - & - & - & - & - & - & - \\
$150^\circ$ & 72 &67 & - & - & - &23 &72 &70 &67 &72 &60 &11 &14 & - & - & - & - & - & - & - \\
$160^\circ$ & 72 &72 &72 & - & - &54 &72 &72 &72 &72 &62 &39 &25 & 3 & 1 & - & - & - & - & - \\
$170^\circ$ & 72 &72 &72 & - &24 &72 &72 &72 &72 &72 &68 &49 &15 & 4 & 1 & - & - & - & - & - \\
$180^\circ$ & 12 &12 &12 & - &11 &12 &12 &12 &12 &12 &12 &12 & 8 & 4 & 1 & - & - & - & - & - \\
\hline
$0^\circ$   & 12 & - & - & - & - & - & - & - & - & - & - & - & - & - & - & - & - & - & - & - \\
$10^\circ$  & 72 & - & - & - & - & - & - & - & - & - & - & - & - & - & - & - & - & - & - & - \\
$20^\circ$  & 60 &17 & - & - & - & - & 1 & 2 & - & - & - & - & - & - & - & 1 & - & 1 & 1 & - \\
$30^\circ$  & 35 & - & - & - & - & - & - & - & - & - & - & - & - & - & - & - & - & - & - & - \\
$150^\circ$ & 37 & - & - & - & - & - & - & - & 5 & - & 1 & - & - & - & - & - & - & - & - & - \\
$160^\circ$ & 69 & - & - & - & - & - & 2 & 1 &10 &13 &15 &10 & 4 & 1 & - & - & - & - & - & - \\
$170^\circ$ & 72 & - & - & - & - & - &25 &16 & 4 & 9 &11 &11 & - & - & - & - & - & - & - & - \\
$180^\circ$ & 12 &12 & - & - & - & - & 6 & 5 & 4 & 2 & 2 & - & - & - & - & - & - & - & - & - \\\hline
\end{tabular}
}
\end{table*}

The numbers of test particles surviving after 1 Myr at each semimajor
axis are given in Table~\ref{table:pevtable}. Shown in
Figure~\ref{fig:survivtimes} are the mean survival times plotted
against semimajor axis for a variety of inclinations. The mean
survival time $\ts$ can be computed by averaging over the results for
the differing longitudes of ascending node and pericentre at fixed
semimajor axis and inclination. The averaging is over 12 test
particles for $i_{\rm tp} = 0^\circ$ and $180^\circ$ and 72 for all
other cases.  Test particles that survive to the end of the
integrations are included with a survival time of 1 Myr, so that the
computed mean survival time is a lower limit in these cases.
Figure~\ref{fig:survivtimes} shows that there is a region of enhanced
stability with $\ts \approx 100$ kyrs -- for all the studied
inclinations -- centred at $\approx 3.5$ au, just beyond the
gravitational reach of the planet but within the critical semimajor
axis and just within the region affected by the secondary.  Most test
particles here (and also beyond this region) suffer a close encounter
with the secondary before ejection or collision with the primary.  The
effect of the secondary star becomes increasingly important with
increasing semimajor axis and $\ts$ falls to $\approx$ 1 kyr. There
is, for the prograde cases, a region of enhanced stability around $12$
au for some inclinations, clearly visible in
Figure~\ref{fig:survivtimes}. This is due to a few particles surviving
here for the full length of the integration.

As stated in the introduction, the mass of star B is uncertain.  By
altering this parameter and rerunning some of the simulations the
importance of the uncertainty can be seen. For the coplanar
configuration of test particles described in this section, changing
the mass of star B by $\sim 20 \%$ results in almost no difference in
global statistical properties, such as average survival times. This
would indicate that the system's dynamics are not significantly
affected by the uncertainty in the secondaries mass.

%--------------------------------------------------------------

%
\begin{figure}
\includegraphics[width=\hsize]{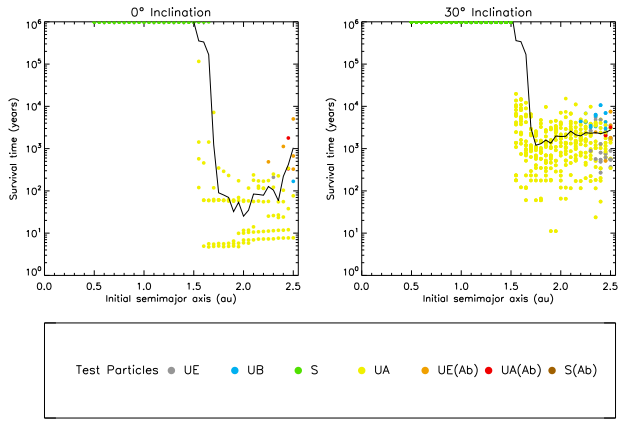}
\caption{\label{fig:starB} The mean survival times (black line) and
fate of individual particles (coloured points) for test particles
around star B. The results are shown for inclinations $0^\circ$ and
$30^\circ$. The fate of each particle is given by the key: UE =
unstable, ejected from the system; UB = unstable, collided with star
B; S = stable; UA = unstable, collided with star A; UE(Ab) = unstable,
ejected from the system after a close encounter with Ab; UA(Ab) =
unstable, collided with star A after a close encounter with Ab; S(Ab)=
stable, but suffered a close encounter with Ab. Note that, in the
retrograde cases, almost all particles are stable for the length of
the integration (1 Myr).}
\end{figure}
\begin{figure}
\includegraphics[width=\hsize]{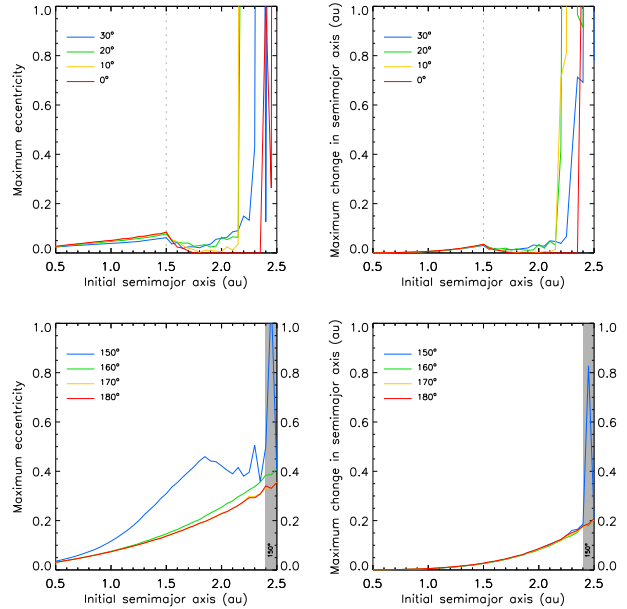}
\caption{\label{fig:starBae} 
  The average maximum eccentricity and maximum change in semimajor
  axis for test particles around star B. This is calculated by
  averaging over longitude at each semimajor axis. The inclinations
  shown in the top panels are: $0^\circ$ (red), $10^\circ$ (yellow),
  $20^\circ$ (green) and $30^\circ$ (blue). The inclinations shown in
  the bottom panels are $180^\circ$ (red), $170^\circ$ (yellow),
  $160^\circ$ (green) and $150^\circ$ (blue).  Also shown as dashed
  vertical lines are the semimajor axis within which all particles
  remain stable. For the $150^\circ$ case the one unstable semimajor
  axis is shaded in grey. }
\end{figure}
\begin{figure}
\includegraphics[width=\hsize]{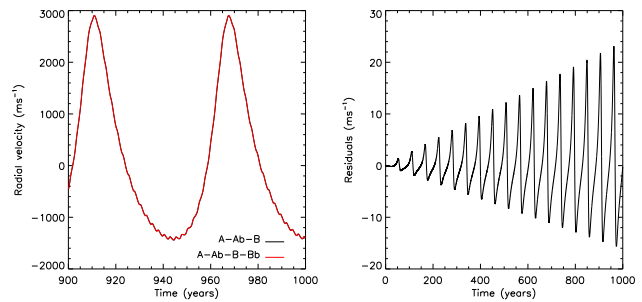}
\caption{\label{fig:rv} 
  The radial velocity curve (left) and residuals (right) for the
  primary star when there is an additional planetary companion to star
  B. The additional planet has a mass of $10 \MJ$. The radial velocity
  curve shows the original system in black and the new two planet
  system in red. The residuals are calculated by subtracting these two
  curves.  When there is no planet around star B the star has an
  increased mass of $M_{\rm{B}} + M_{\rm{Bb}}$, where $M_{\rm{Bb}}$ is
  the mass of the additional planet to be added to the later
  simulation.}
\end{figure}

\section{Planets and Asteroid Belts around the Secondary}

\label{sec:b}

Here, we investigate whether planets could exist around star B.  This
possibility has not been investigated before for this system.  The
critical semimajor axis for stability, as defined by Holman \& Wiegert
(1999), is $1.9 \pm 1.0$ au. To investigate further, test particles
are set up from $0.5$ to $2.5$ au in steps of $0.05$ au for
inclinations $0^\circ$ to $30^\circ$ and $150^\circ$ to $180^\circ$ in
steps of $10^\circ$. The particles are on initially circular orbits
again, and spaced in longitude of perihelion and ascending node by
$60^\circ$ as before.  The results of two of the prograde cases are
shown in Figure~\ref{fig:starB}. Prograde test particles are not
stable beyond $1.5$ au. In the case of $10^\circ$, $20^\circ$ and
$30^\circ$ inclinations, the unstable test particles generally survive
at least ten times longer than those in the $0^\circ$ case. The
unstable test particles within about $2$ au all collide with the
primary, whilst those beyond this point have a range of fates.  Holman
\& Wiegert's critical semimajor axis does not match up as well as in
Section~\ref{sec:tpib}, but still agrees within the rather large
uncertainty.  In the retrograde cases, almost all the test particles
are stable, with the exception of 7 particles at $2.45$ au in the
$150^\circ$ case.  Integrating more distant particles shows that the
retrograde stability reaches out to about $3.5$ au.

The stability of the test particles can be further investigated by
plotting the evolution of their eccentricity and semimajor axis, as
shown in Figure~\ref{fig:starBae}.  For all the simulations, there is
not much change in the inclination of the test particles. The
variation in eccentricity and semimajor axis of stable particles
increases as the initial distance from star B increases, and is
similar for all the prograde cases.  However, the variation is much
larger for the (more stable) retrograde cases. When $i_{\rm tp} =
150^\circ$, the variations are no longer smoothly increasing and show
abrupt jumps, indicating that this case may not be long term stable.

As test particles can remain stable around the secondary, this raises
the question of whether a massive planet could also persist here. So,
it is interesting to consider whether this could be detectable in the
radial velocity curve of the primary star.  To investigate this, the
simple case of a $10$ Jupiter-mass planet in a coplanar, circular
orbit with initial semimajor axis $0.5$ au and initial longitude
$0^\circ$ is integrated. The orbital elements of the secondary are
adjusted so that the centre of mass of it and its putative planet
orbits the primary with the elements shown in Table 1 for star B
alone.  As expected from the test particle results, the planet remains
stable for the full Myr length of the integration and shows very
little variation in its orbit.  However, the 1 Myr integrations here
may overestimate the extent of the stability zone for planets, as
instabilities can appear even after 100 Myr of apparent stability
(Jones, Sleep \& Chambers 2001).  To calculate the radial velocity
curve, the system is assumed to be at zero inclination relative to the
line of sight ($i = 90^\circ$).  Figure~\ref{fig:rv} shows the radial
velocity curve of the primary, together with the residuals with
respect to the case with no extra planet. There is no detectable
signal with the period of the extra planet. The small variations
shown, which have the period of the binary, amount to 25 ms$^{-1}$
over 1000 yr timespans, would be undetectable.

%
%--------------------------------------------------------------

\section{Edgeworth-Kuiper Belt Analogues}

\label{sec:kb}

The final region remaining to be studied is that exterior to the
binary. This may host particles analogous to the Edgeworth-Kuiper belt
in our own Solar system. There is observational evidence for the
existence of extrasolar Edgeworth-Kuiper belts from infrared imaging
of dusty discs around other stars (e.g., Wyatt 2003, Greaves et al.
2004), so the longevity of such a structure around $\gamma$ Cephei is
worth investigation.

Numerical integrations of test particles around binary systems suggest
that they will remain stable in this region (e.g., Harrington
1977). In addition to the criterion in Section~\ref{sec:tpib}, Holman
\& Wiegert (1999) also provide an empirical rule-of-thumb to determine
prograde test particle stability exterior to the binary. Here, the
critical semimajor axis $a_{\rm crit}$ is that beyond which almost all
test particles survive for at least $10^4$ binary periods. Using
eq. (3) of their paper, we find that $a_{\rm crit} = 64 \pm 2$ au.
However, Holman \& Wiegert (1999) caution that there can sometimes be
islands of instability beyond the critical radius associated with
$n$:1 MMRs. There is an older criterion due to Harrington (1977), who
also considers the case of retrograde test particles. Harrington's
equation suggests that prograde test particles are stable for $a \gta
56$ au and retrograde stable for $a \gta 45$ au.  By stability,
Harrington means that the particles show bounded motion with no
secular or large periodic variations in their elements over his -- by
nowadays standards -- very small integration timespans.
\begin{figure}
  \includegraphics[width=\hsize]{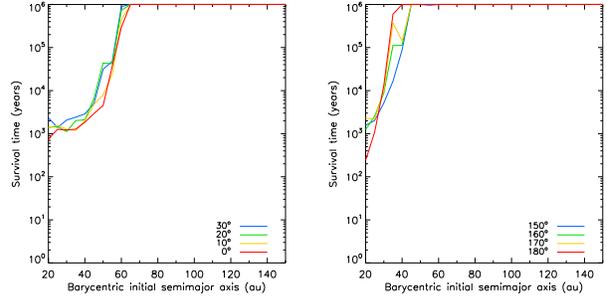}
\caption{\label{fig:outer} 
  Average survival time of test particles exterior to the binary as a
  function of barycentric initial semimajor axis. In the left hand
  panel the starting inclinations are: $0^\circ$ (red), $10^\circ$
  (yellow), $20^\circ$ (green) and $30^\circ$ (blue). In the right
  hand panel the starting inclinations are: $180^\circ$ (red),
  $170^\circ$ (yellow), $160^\circ$ (green) and $150^\circ$ (blue).  }
\end{figure}
Using the same method as in Sections~\ref{sec:tpip} and
\ref{sec:tpib}, test particles are set up with inclinations $0$ to
$30^\circ$ and $150^\circ$ to $180^\circ$ in the (barycentric) region
from $20$ to $150$ au in steps of 5 au. The mean survival times are
shown in Figure~\ref{fig:outer}. The prograde test particles exhibit a
sharp cut-off, with those beyond $65$ au being stable, independent of
the starting inclination.  This figure also shows that the retrograde
test particles survive to the end of the integration for starting
semimajor axes beyond 40 au for $180^\circ$ inclination, 45 au for
$170^\circ$ and $160^\circ$ inclinations and 60 au for $150^\circ$
inclination. The range of initial conditions for which retrograde
particles survive is larger than for prograde (e.g., Harrington 1977).
For all inclinations, unstable particles from about $20$ to $40$ au
are removed quickly and generally are ejected after a close encounter
with the secondary. Unstable particles from about $40$ to $65$ au are
less rapidly removed and tend to collide with star A or be ejected
from the system, without experiencing a prior close encounter. We see
that both Harrington's (1977) and Holman \& Wiegert's (1999) stability
criteria seem to give a reasonable description of the results of our
simulations.

\begin{figure}
\includegraphics[width=\hsize]{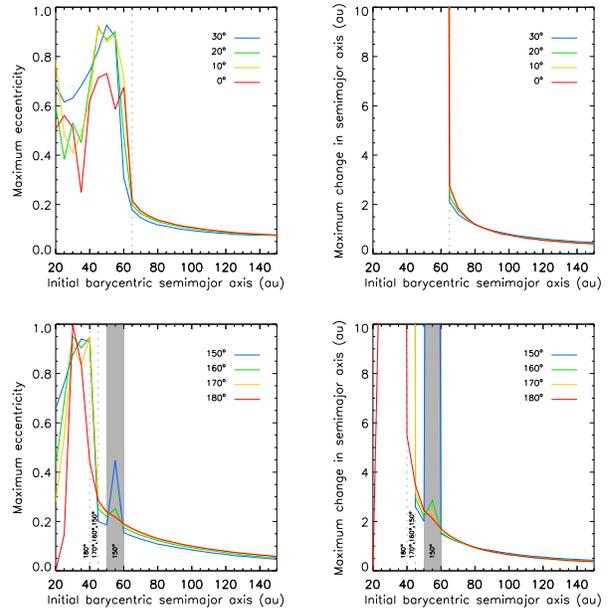}
\caption{\label{fig:baryboth} 
  The average maximum eccentricity and maximum change in semimajor axis
  of each test particle in the region exterior to the binary, as a
  function of starting semimajor axis. The inclinations shown are as
  for Figure~\ref{fig:outer}, and the averages calculated as for
  Figure~\ref{fig:starBae}. The boundaries between the unstable and
  stable regions are shown as dashed grey lines. An isolated unstable
  region is shown as a shaded area for the $150^\circ$ case, as in
  Figure~\ref{fig:starBae}.}
\end{figure}

The average maximum eccentricity and change in semimajor axis is shown
in Figure~\ref{fig:baryboth} as a function of initial semimajor axis
for all eight inclinations. All the test particles show a small
secular variation in their orbital elements.

\begin{figure}
\includegraphics[width=\hsize]{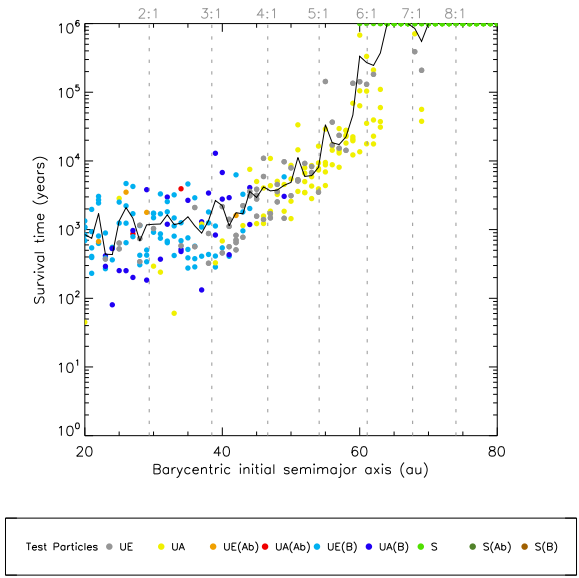}
\caption{\label{fig:14} 
  The average survival times (black line) and fate of individual
  particles (coloured points) for test particles in the region
  exterior to the binary. This is for an inclination of $0^\circ$ and
  with a resolution of $1$ au in semimajor axis. Only the particles
  out to $80$ au are shown, as the simulation is stable beyond here.
  The locations of the $n$:1 MMRs are shown as dashed grey lines and
  labelled above the plot. The colour coded fates of test particles
  are as in Figure~\ref{fig:inner}.}
\end{figure}

The Edgeworth-Kuiper belt in our Solar system shows some structure due
to locations of MMRs with the giant planets (e.g., Luu \& Jewitt
2002). However, the results in Figure~\ref{fig:outer} do not indicate
any such resonant features.  This is most likely due to the coarse
semimajor axis grid employed. To investigate this, the coplanar case
was re-run, but now with a spacing in semimajor axis of $1$ au. The
mean survival times and fates of individual particles for this
simulation are shown in Figure~\ref{fig:14}. The first location where
all test particles survive is now $64$ au, agreeing exactly with
Holman \& Wiegert's value of $a_{\rm crit}$. There is then an unstable
band from $67$ to $70$ au that seems to match up to the 7:1 MMR. The
plot gives no evidence for any other resonant effects beyond this
location.  This is understandable since the locations of the other
major resonances are outside the region of Hill stability (for
example, the 3:2 resonance that is important in our own Solar system),
leaving only those of very high order to affect stable test particles.
The distinct difference in fates of test particles is obvious in this
plot, with the blue coloured points, indicating close encounters with
star B, not occurring beyond about $45$ au. At this point, the test
particles are yellow or grey, indicating either ejection or collision
with star A without undergoing any close encounter.

%--------------------------------------------------------------

\section{Conclusions}

In this paper, we have carried out a suite of test particle
integrations for the $\gamma$ Cephei system, as a guide to possible
locations for additional planets. For test particles in the plane of
the binary and planet, there are three zones of stability for 1 Myr
timespans at least. These are [1] the region interior to the planet,
which is stable within the bands $1.2$ and $1.3$ au, $1.05$ to $1.15$
au, and $0.75$ to at least $0.5$ au, [2] the region around star B from
$1.5$ au into at least $0.5$ au and [3] the region around both stars
extending out from about $65$ au. These can be used to constrain
possible locations of additional planetary companions.

The region interior to the planet has a complicated structure.  Low
order resonances, such as the 4:1, 3:1 and 5:2, mark transitions
between stable and unstable regimes. There is a secular resonance
located at $0.8$ au from the primary that also plays a role in the
dynamics of the test particles.  The results for this region match up
well with the previous work of Dvorak et al. (2003) despite the
differences in the parameters of the system and the methods used. This
implies that the slight improvements in the observations of the system
have not significantly altered its dynamical characteristics.  One
difference though is that, unlike Dvorak et al. (2003), we find that
test particles close to the 3:1 resonance are unstable, just as for
the asteroid belt in the Solar System. In agreement with Dvorak et
al. (2003), we find that small planets interior to $\gamma$ Cephei Ab
are long-lived.  Another as yet unconsidered possibility seen from the
results here is the existence of an asteroid belt interior to the
giant planet. However, the stability seen for test particles in this
region disappears when $\gamma$ Cephei Ab is inclined in the same
plane as the test particles. Now, the test particles are rapidly
driven to inclinations which are subject to the Kozai instability.

The region interior to the binary and exterior to the planet is
unpromising. It seems unlikely that any prograde planet or asteroid
can remain stable between $\gamma$ Cephei Ab and the binary, although
retrograde test particles in this region are more long-lived.  The
region around the secondary is stable out to $1.5$ au for test
particles. We have shown that planets up to 10 Jupiter-masses in
circular orbits at $0.5$ au can survive for at least 1 Myr.  This
seems to be a promising place for additional planets to reside, in a
similar manner to satellites about a planet around a single star.

In the final region, that exterior to the binary, test particles are
also stable. Although planets would be able to reside out here, the
existence of an Edgeworth-Kuiper belt structure is also a possibility.
Once again the retrograde particles are more stable than those that
are prograde. Whilst in every region studied this is true, such
objects are perhaps unlikely, if the origin of the system was a common
disk.  If particles are captured from elsewhere, then this may become
a possibility.  Retrograde asteroids and comets certainly exist in the
Solar system.

Since the survival of additional planets in the system has been shown
to be a possibility, it is interesting to consider their habitability.
There are a number of estimates of the habitable zone around star A in
the literature. For example, Dvorak et al. (2003) place it at $1.0$ to
$2.2$ au, in which case habitable planets could exist at the very edge
of the zone.  However, Haghighipour (2005) places the habitable zone
at $3.1$ to $3.8$ au, while Jones et al. (2005) place it at $2.07$ to
$4.17$ au.  These different locations reflect differences in the
criteria for the location of the habitable zone or the assumed stellar
luminosity and effective temperature.  The zone $2.07$ to $4.17$ au is
unstable for all except retrograde test particles. So, this would
permit the possibility of a habitable planet only if retrograde. The
detection of any such small planet is challenging, given the small (of
the order of a few ms$^{-1}$) radial velocity signatures that they
cause. In addition, a retrograde planet appears in radial velocity
curves merely as a prograde one with a $180^\circ$ phase difference.
Although retrograde objects -- especially planets -- are thought to be
unlikely, it has been shown that giant planets in binary systems can
end up with retrograde orbits after close encounters within a
planetary system (Marzari et al. 2005).

A more interesting possibility is that of a habitable planet around
star B, since it has been shown that planets can survive
here. Theories of planetary formation do not preclude this possibility
(e.g., Armitage, Clarke \& Palla 2003). The habitable zone for an M
dwarf extends out to about 0.3 au (Kasting, Whitmire \& Reynolds
1993), which is within the stable zone found here. The large primary
star nearby might also act as a `shield' from comets and asteroids
similar to Jupiter for the Earth. As seen in the results in
Sections~\ref{sec:b} and \ref{sec:kb}, very few test particles collide
with the secondary star once the region interior to the binary has
been rapidly cleared. Edgeworth-Kuiper belt objects that are perturbed
into the inner regions of the system are also likely to suffer
encounters with the primary, thus leaving the secondary and its
environs largely unscathed.  

The detection of such a habitable planet around star B has,
unfortunately, been shown to be virtually impossible from the radial
velocity signature of star A alone. Should the secondary be resolved
this would change, and any companion giant planet would be easily
detectable.  The region around star B, none the less, remains as a
promising place for the existence of a habitable planet in this
system.

\section*{Acknowledgments}
PEV acknowledges financial support from the Particle Physics and
Astronomy Research Council. We thank the referee for his helpful
report.

%--------------------------------------------------------------

\label{lastpage}


\begin{thebibliography}{}

\bibitem[Affer et al.(2005)]{Af05}
Affer L., Micela G., Morek T., Sanz-Forcada J., Favata F. 2005, A\&A,
433, 647
%
\bibitem[Armitage, Clarke \& Palla (2003)]{Ar03}
Armitage P.~J., Clarke C.~J., Palla F. 2003, MNRAS, 342, 1139
%
\bibitem[Chambers (1999)]{Ch99}
Chambers J. 1999, MNRAS, 304, 793

\bibitem[Dvorak et al.(2003)]{Dv03}
Dvorak R., Pilat-Lohinger E., Funk B., Freistetter F. 2003, A\&A, 398, L1

\bibitem[Dvorak et al.(2004)]{Dv04} Dvorak, R., 
Pilat-Lohinger, E., Bois, E., Funk, B., Freistetter, F.,
Kiseleva-Eggleton, L. 2004, Revista Mexicana de Astronomia y Astrofisica 
Conference Series, 21, 222 

\bibitem[Greaves et al.(2004)]{2004MNRAS.351L..54G} Greaves J.~S., Wyatt 
M.~C., Holland W.~S., Dent W.~R.~F. 2004, MNRAS, 351, L54 
 
\bibitem[Ford et al.(2000)]{Fo00} Ford, E.~B., Kozinsky, B., 
Rasio, F.~A. 2000, ApJ, 535, 385 

\bibitem[Fuhrmann (2004)]{Fu04}
Fuhrmann K., 2004, Astron. Nach., 325, 3

\bibitem[Griffin et al.(2002)]{2002Obs...122...90G} Griffin, R.~F., 
Carquillat, J.-M., Ginestet, N. 2002, The Observatory, 122, 90 
 
\bibitem[Hatzes et al.(2003)]{Ha03}
Hatzes A.P., Cochran W.D., Endl M., McArthur B., Paulson D.~B., 
Walker G.A.H., Campbell B., Yang S. 2003, ApJ, 599, 1383

\bibitem[Haghighipour (2005)]{Ha05}
Haghighipour N. 2005, ApJ, submitted (astro-ph/0509659)

\bibitem[Harrington (1977)]{Ha77}
Harrington R.S. 1977, AJ, 82, 753

\bibitem[Holman \& Wiegert(1999)]{1999AJ....117..621H} Holman, M.~J., \& 
Wiegert, P.~A. 1999, AJ, 117, 621 

\bibitem[Jones, Sleep \& Chambers(2001)]{Jo01} Jones B.~W., Sleep P.~N.,
Chambers, J.~E. 2001, A\&A, 366, 254 

\bibitem[Jones et al.(2005)]{Jo05} Jones B.~W., Underwood D.~R., 
Sleep P.~N. 2005, ApJ, 622, 1091 

\bibitem[Kasting et al.(1993)]{Ka93} Kasting J.~F., 
Whitmire D.~P., Reynolds R.~T. 1993, Icarus, 101, 108 
 
\bibitem[Kozai(1962)]{Ko62} Kozai, Y. 1962, AJ, 67, 591 

\bibitem[Luu \& Jewitt(2002)]{2002ARA&A..40...63L} Luu J.~X., Jewitt 
D.~C. 2002, ARAA, 40, 63  

\bibitem[Marzari et al.(2005)]{Ma05} Marzari F., Weidenschilling S. J.,
Barbieri M., Granata V. 2005, ApJ, 618, 502

\bibitem[Murray \& Dermott(2000)]{Mu00} Murray C.~D., 
Dermott S.~F. 2000, Solar System Dynamics, Cambridge University Press, 
chap. 7

\bibitem[Press et al.(1999)]{Pr99} Press W.H., Flannery B.P.,
Teukolsky S.A., Vetterling W.T. 1999, Numerical Recipes, Cambridge
University Press, Cambridge

\bibitem[Smart(1977)]{smart} Smart W.M. 1977, Spherical Astronomy,
chap 14

\bibitem[Wyatt et al.(2003)]{2003EM&P...92..423W} Wyatt M.~C., Holland 
W.~S., Greaves J.~S., Dent W.~R.~F. 2003, Earth Moon and Planets, 92, 
423 

\bibitem[Wyatt et al.(2005)]{2005ApJ...620..492W} Wyatt M.~C., Greaves 
J.~S., Dent W.~R.~F., Coulson I.~M. 2005, ApJ, 620, 492 
  
\end{thebibliography}
\end{document}